\documentclass[12pt]{article}
\usepackage{epsfig,epsf,xcolor,verbatim}
\usepackage{amsmath,amssymb,float,tikz}
\usepackage{graphicx}

\begin{document}


\begin{center}
{\Large\bf Lens rigidity  in scattering by non-trapping obstacles}
\end{center}

\begin{center}
{\sc Luchezar Stoyanov}
\end{center}

\newcommand{\R}{{\sf I\hspace{-.15em}R}}
\newcommand{\SSS}{{\sf I\hspace{-.4em}S}}
\def\C{{\bf C}}
\def\e{\emptyset}
\def\ds{\partial S}
\def\cq{\overline{Q}}
\def\sn{{\bf S}^{n-1}}
\def\sd{{\bf S}^{d-1}}

\def\hg{\Gamma}
\def\ssn{{\sn}\times {\sn}}
\def\do{\partial \Omega}
\def\dk{\partial K}
\def\dl{\partial L}
\def\tts{\tilde{S}}
\def\ttp{\tilde{p}}
\def\tu{\tilde{U}}
\def\hs{\hat{S}}
\def\bS{{\bf S}}
\def\hp{\hat{p}}
\def\dr{\frac{\partial r}{\partial x_1}}
\def\ll{{\cal L}}
\def\hm{\hat{M}}
\def\tM{\widetilde{M}}
\def\pp{{\cal P}}
\def\tt{{\cal T}}
\def\ds{\partial S}
\def\ss{{\cal S}}
\def\vv{{\cal V}}
\def\aa{{\cal A}}
\def\bb{{\cal B}}
\def\nn{{\cal N}}
\def\dd{{\cal D}}
\def\ee{{\cal E}}
\def\uu{{\cal U}}
\def\rr{{\cal R}}
\def\kk{{\cal K}}
\def\cc{C_0^{\infty}}
\def\ot{(\omega,\theta)}
\def\oto{(\omega_0,\theta_0)}
\def\toto{(\tilde{\omega}_0, \tilde{\theta}_0)}
\def\ts{\tilde{\sigma}_0}
\def\tx{\tilde{x}}
\def\txi{\tilde{\xi}}
\def\got{\gamma(\omega,\theta)}
\def\ggot{\gamma'(\omega,\theta)}
\def\omo{(\omega,\omega)}
\def\du{\partial u_i}
\def\dfu{\frac{\partial f}{\partial u_i}}
\def\dou{\frac{\partial \omega}{\partial u_i}}
\def\dttu{\frac{\partial \theta}{\partial u_i}}
\def\ooo{(\omega_0,-\omega_0)}
\def\otp{(\omega',\theta')}
\def\oo{{\cal O}}
\def\pr{{\rm pr}}
\def\cK{\hat{K}}
\def\cL{\hat{L}}
\def\ff{{\cal F}}
\def\fk{{\cal F}^{(K)}}
\def\fl{{\cal F}^{(L)}}
\def\kkr{{\cal K}^{\mbox{reg}}}
\def\kkro{{\cal K}_0^{\mbox{reg}}}
\def\trapk{\mbox{\rm Trap}(\Omega_K)}
\def\trapl{\mbox{\rm Trap}(\Omega_L)}
\def\sock{S^*(\stackrel{\circ}{\Omega}_{\cK})}
\def\oock{\stackrel{\circ}{\Omega}_{\cK}}
\def\socl{S^*(\stackrel{\circ}{\Omega}_{\cL})}
\def\oocl{\stackrel{\circ}{\Omega}_{\cL}}
\def\sok{S_b^*(\Omega_K)\setminus \trapk}
\def\sol{S_b^*(\Omega_L)\setminus \trapl}
\def\sbok{S_b^*(\Omega_K)}
\def\sbol{S_b^*(\Omega_L)}
\def\hr{\hat{\rho}}
\def\G{{\cal G}}
\def\tg{\tilde{\gamma}}
\def\Vol{\mbox{Vol}}
\def\dkso{\partial K^{(\infty)}}
\def\dlso{\partial L^{(\infty)}}
\def\gg{{\cal g}}
\def\sl{{\cal SL}}
\def\tkn{\mbox{\rm Trap}^{(n)}(\dk)}
\def\tln{\mbox{\rm Trap}^{(n)}(\dl)}
\def\wn{{\cal WN}}
\def\i{{\bf i}}
\def\te{{\cal T}^{(ext)}}
\def\ndk{{\cal N}_b^*(\dk)}
\def\ndl{{\cal N}_b^*(\dl)}
\def\dr{\frac{\partial r}{\partial z_1}}
\def\endofproof{{\rule{6pt}{6pt}}}
\def\Box{\endofproof}
\def\su{S^*(\R^n\setminus U)}
\def\gk{\gamma_K}
\def\gl{\gamma_L}
\def\la{\left\langle}
\def\ra{\right\rangle}
\def\kfin{\kk_0^{({\mbox{\footnotesize\rm fin}})}} 
\def\dt{\dot{T}}
\def\ep{\epsilon}
\def\kfi{\kk^{({\mbox{\footnotesize\rm fin}})}} 
\def\stk{\Sigma_3^{(K)}}
\def\stl{\Sigma_3^{(L)}}
\def\SU{S^*(\R^n\setminus U)}
\def\tkm{\tt_k^{(m)}}
\def\ukm{U_k^{(m)}}
\def\Pkm{\Psi_k^{(m)}}
\def\nkm{N_k^{(m)}}
\def\di{\displaystyle}
\def\nk{N^{(K)}}
\def\nl{N^{(L)}}
\def\mk{M^{(K)}}
\def\ml{M^{(L)}}
\def\ep{\epsilon}
\def\Gk{G^k}
\def\uk{U^{(K)}}
\def\dist{\mbox{\rm dist}}
\def\diam{\mbox{\rm diam}}
\def\con{\mbox{\rm const}}
\def\bs{\bigskip}
\def\ms{\medskip}
\def\SK{S^*_{K}(S_0)}
\def\SL{S^*_{L}(S_0)}
\def\id{\mbox{\rm id}}
\def\tX{\widetilde{X}}
\def\tY{\widetilde{Y}}
\def\tF{\widetilde{F}}
\def\tsigma{\tilde{\sigma}}
\def\hY{\widehat{Y}}
\def\tv{\tilde{v}}
\def\rank{\mbox{\rm rank}}
\def\sl{{\mathcal SL}}
\def\ot{(\omega,\theta)}
\def\oto{(\omega_0,\theta_0)}
\def\tOm{\widetilde{\Omega}}

\def\te{\tilde{e}}
\def\tx{\tilde{x}}
\def\ty{\tilde{y}}
\def\grad{\; \mbox{\rm grad} \,}
\def\spp{\vspace{5pt}\noindent}
\newtheorem{definition}{Definition}
\newtheorem{construction}{Construction}
\newtheorem{example}{Example}
\newtheorem{lemma}{Lemma}
\newtheorem{theorem}{Theorem}
\newtheorem{corollary}{Corollary}
\newtheorem{proposition}{Proposition}

\footnotesize

\noindent
{\bf Abstract.} We prove that if two non-trapping obstacles in $\R^n$ satisfy some rather weak non-degeneracy conditions
and the scattering rays in their exteriors have  (almost) the same travelling times or (almost) the same scattering length spectrum, 
then they coincide.

\medskip

\medskip 
\noindent
{\bf Mathematics Subject Classification:} 37D20, 37D40, 53D25, 58J50

\noindent
{\bf Keywords:}
scattering by obstacles, billiard trajectory,  travelling time, scattering length spectrum
\normalsize

\def\do{\partial \Omega}

\section{Introduction}\label{sec1}
\renewcommand{\theequation}{\arabic{section}.\arabic{equation}}
\setcounter{equation}{0}

We consider scattering by obstacles $K$ in $\R^n$, $n \geq 2$. Here $K$ is compact subset of ${\R}^n$ 
with a $C^{3}$ boundary $\partial K$ such that  $\Omega_K = \overline{{\R}^n\setminus K}$ is connected. 
By a {\it scattering ray} in  $\Omega_K$ we mean an unbounded in both directions generalized geodesic 
(in the sense of Melrose and  Sj\"ostrand \cite{MS1}, \cite{MS2}). As is well-known (see Sect. 2 for more information),
most of these scattering rays are billiard trajectories with finitely many reflection points at $\dk$.

Given an obstacle $K$ in $\R^n$, consider a large ball $M$  containing $K$ in its interior, and let $\bS_0 = \partial M$
be its boundary sphere.  Then
$$\Omega_K = \overline{M\setminus K}$$
is a compact subset of $M$ with smooth boundary $\do_K = \partial M \cup \dk$. Denote by 
$\fk_t$ ($ t\in\R$) the {\it generalized geodesic flow} on the co-sphere bundle $\bS^*(\overline{\R^n\setminus K})$.
Since most trajectories of $\fk_t$ are billiard trajectories, we will sometimes call it the {\it billiard flow} in
the exterior of $K$.
For any $q\in \do_K$ let $\nu(q)$ the {\it unit normal} to $\do_K$ pointing into the interior of $\Omega_K$, and let
$$\bS^*_+(\bS_0 ) = \{ x = (q,v) : q \in \bS_0 \:, \: v \in \sn\: ,\:  \la v, \nu(q) \ra \geq 0\} .$$
Given $x \in  \bS^*_+(\bS_0)$, define the {\it travelling time} $t_K(x)\geq 0$ as the maximal number (or $\infty$) 
such that $\pr_1(\fk_t(x))$ is in the interior of $\Omega_K$ for all $0 < t < t_K(x)$, where $\pr_1(p,w) = p$. 
For $x = (q,v) \in \bS^*_+(\bS_0)$ with $\la \nu(q),v\ra = 0$ set $t(x) = 0$.

It is a natural problem to try to recover information about the obstacle $K$ from the {\it travelling times} $t_K(x)$,
$ x \in \bS^*_+(\bS_0)$. Similar problems have been actively considered in 
Riemannian geometry  -- see \cite{SU}, \cite{SUV} and the references there for some 
general information. In scattering by obstacles this kind of problems have been studied  as well -- 
see \cite{LP}, \cite{M} and \cite{PS} for general information, and also the more recent papers \cite{St2}, \cite{St3},  
\cite{NS1}, \cite{NS2} and the references there. In particular, it has been established that for 
some classes of obstacles $K$ knowing all (or almost all with respect to the Lebesgue measure) travelling times 
$t(x) = t_K(x)$ completely determines $K$. This is so, for example, in the class 
of obstacles $K$ that are finite disjoint unions of strictly convex domains with smooth boundaries (\cite{NS1}). 
It has been an open problem so far whether a similar result holds for general  non-trapping obstacles. 
Here we prove that this is the case under some mild non-degeneracy assumptions about the obstacle. 

A natural impediment in trying to recover information about the obstacle $K$ from its travelling times is the set
$\trapk$ of trapped points. A point  $x = (q,v)\in  \bS(\Omega)$ is called {\it trapped} if either  its forward billiard trajectory 
$$\gk^+(x) = \{\pr_1(\fk_t(x)) : t \geq 0\} $$ 
or  its backward trajectory $\gk^- (q,v) = \gk^+(q, -v)$ is infinitely long. That is, either the billiard 
trajectory in the exterior of $K$ issued from $q$ in the direction of $v$ is bounded (contained entirely in $M$) or the 
one issued from $q$ in the direction of $-v$ is bounded. The obstacle $K$ is called {\it non-trapping} if $\trapk = \e$.

It is known (see [\cite{LP} or Proposition 2.3 in \cite{St1}) that $\trapk \cap \bS^*_+(\bS_0)$ has Lebesgue
measure zero in $\bS^*_+(\bS_0)$. However, as an example of M. Livshits shows  (see  Figure 1),  in general  
the set of trapped points $x\in \bS(\Omega_K)$  may contain a non-trivial open set, and then the recovery of the  obstacle 
from travelling times is impossible. Similar examples in higher dimensions were constructed in \cite{NS3}.

Denote by $\kk$ be the {\it class of obstacles with the following property}: for each 
$(x,\xi)\in \dt^* (\partial K) = T^*(\dk) \setminus \{0\}$ if the curvature 
of $\partial K$ at $x$ vanishes of infinite order in direction $\xi$, then all points $(y,\eta)$ sufficiently close to $(x,\xi)$ 
are diffractive points, i.e. $\dk$ is convex at $y$ in the direction of $\eta$ (see e.g. \cite{H} or Ch.1 in \cite{PS} for the
formal definition of a diffractive point). 
Given $\sigma = (x,\xi) \in \bS^*_+(\bS_0)$, we  will say that the trajectory
$\gamma_K(\sigma)$ is {\it regular} (or {\it non-degenerate}) if for every $t >> 0$ the differential of the map
$\R^n \ni y \mapsto \mbox{pr}_2(\fk_t(y,\xi))\in \sn$ has maximal rank $n-1$ at $y=x$, and also
the differential of the map $\sn \ni \eta \mapsto \mbox{pr}_1(\fk_t(x,\eta))\in \R^n$ has maximal rank $n-1$ at $\eta = \xi$.

Let $\kk_0$ the {\it class of all obstacles} $K\in\kk$  such that $\dk$ does not contain  non-trivial open flat subsets and  
$\gamma_K(x,u)$ is a regular simply reflecting ray for almost all $(x,u) \in \bS^*_+(\bS_0)$  such that $\gamma(x,u)\cap \dk\neq \e$. 
Using the technique developed in Ch. 3 in \cite{PS} one can show that   $\kk_0$ is of second Baire category  in $\kk$ with respect 
to the $C^\infty$  Whitney topology in $\kk$. This means that generic obstacles $K$ in $\R^n$ belong to the class $\kk_0$.

Our main result concerns obstacles $K$ satisfying  the following  conditions:

\ms

\noindent
{\bf (A0):} $K \in \kk_0$,

\ms

\noindent
{\bf (A1):} {\it The set $\trapk \cap \bS^*_+(\bS_0)$ is totally disconnected.}

\ms 
 
\noindent
{\bf (A2):}  {\it There exists a finite or countable family $\{C_i\} = \{C^{(K)}_i\}$ of codimension two 

\hspace{0.35cm} $C^1$ submanifolds of $\bS^*_+(\bS_0)\setminus \trapk$  which is locally finite in  $\bS^*_+(\bS_0)\setminus \trapk$

\hspace{0.35cm}  and  such that for every $\rho \in \bS^*_+(\bS_0)\setminus (\trapk \cup \cup_i C_i)$ the billiard trajectory

\hspace{0.35cm}  $\gamma_K(\rho)$ has no conjugate points that both belong to $\dk$.}

\bs

Clearly (A0) is just a very weak non-degeneracy condition.
The conditions in  (A1) imply that the  set of  trapped points is relatively small; as Livshits' example shows without some condition
of this kind complete recovery of the obstacle from travelling times is impossible. The condition (A2) is a bit more subtle,
however it appears to be rather general. In fact it might be generic\footnote{Indeed, locally at least, it
is trivial to `destroy' a pair of conjugate points both lying on $\dk$, simply by perturbing slightly $\dk$ near one of
these points.}, however we do not have a formal proof of this. We refer the reader to \cite{dC} for general information
about Jacobi fields and conjugate points in Riemannian geometry and in particular to \cite{W} for these concepts in the case
of billiard flows\footnote{Albeit in the simpler two-dimensional case, however the higher dimensional case is similar.}.

In this paper we prove the following.

\bs

\noindent
{\bf Theorem 1.1.} { \it Let $K$ and $L$ be two obstacles in $M$ with $C^k$ boundaries ($k\geq 3$)
so that the billiard flow in $\Omega_K$ and the one  in $\Omega_L$ satisfy the conditions 
{\rm (A0), (A1)} and {\rm (A2)}.  Assume that $t_K(x) = t_L(x)$ for almost all $x\in \bS^*_+(\bS_0)$. Then $K = L$.} 

\bs

A similar result holds replacing the assumption about travelling times by a condition about sojourn times of scattering rays
-- see Sect. 2 for details.

The rest of the paper is devoted to the proof of this theorem. The proof in Sect. 3 below is derived from some results 
and arguments in \cite{St3} and \cite{NS1}. Some of these are described in Sect. 2.

\bs

\begin{tikzpicture}[xscale=1.84,yscale=0.77] 
  \draw[blue] (-1.5,3) arc (0:180:2cm and 2cm); 
  \draw (0,1) arc (-30:210:4cm and 4cm); 
  \draw (-3.8,2.3) arc (0:180:0.5cm and 0.7cm); 
  \draw (-2.1,2.3) arc (0:180:0.5cm and 0.7cm); 
  \draw (-3,5) node[anchor=south west] {$E$}; 
  \draw (-4.4,3) node[anchor=north west] {$f_1$};
  \draw (-2.4,3) node[anchor=north east] {$f_2$};
    \draw (-5.9,3.3) node[anchor=north west] {$P$};
  \draw (-1.1,3.3) node[anchor=north east] {$Q$};

\draw (-6.92,1) .. controls (-5.6,-1) and (-3.7,-0.7) .. (-3.8,2.29);
\draw (-3.1,2.29) .. controls (-3.2,-1) and (-1,-0.7) .. (0,1);
\draw[thick] (-2.1,2.29) .. controls (-2,1.2) and (-1.5,1.5) .. (-1.5,3);
\draw[thick] (-5.5,3.1) .. controls (-5.5,1.1) and (-4.9,1.4) .. (-4.8,2.4);

\draw[thick] (-5.5,3) circle (0.02cm);
\draw[thick] (-4.3,3) circle (0.02cm);
\draw[thick] (-2.6,3) circle (0.02cm);
\draw[thick] (-1.5,3) circle (0.02cm);

\end{tikzpicture}

\bigskip

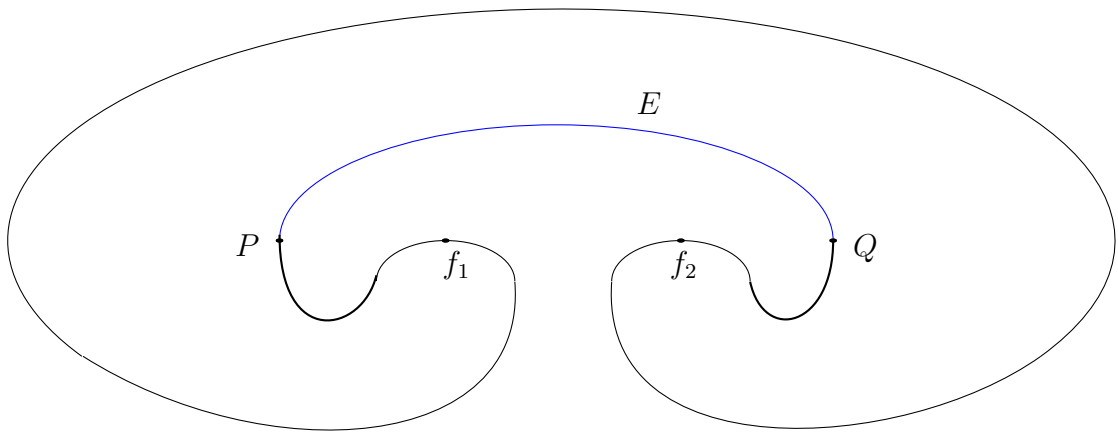
\centerline{Figure 1: Livshits' Example, adapted from Ch. 5 of \cite{M}}

\smallskip

\noindent
{\footnotesize{Here $K$ is an obstacle in $\R^2$ bounded by the closed curve whose part $E$ is half an ellipse with end points $P$ and $Q$
and foci $f_1$ and $f_2$.  Any scattering trajectory entering the area inside the ellipse 
between the foci $f_1$ and $f_2$, will  reflect at $E$ and then go out between the foci again. So, no scattering ray `coming from infinity'
can have a common point with the bold lines from $P$ to $f_1$ and from $Q$ to  $f_2$.}}

\section{Some useful results from previous papers}
\renewcommand{\theequation}{\arabic{section}.\arabic{equation}}
\setcounter{equation}{0}

Here we describe some  previous results  which will be essentially used
in the proof of Theorem 1.1. We also state a more general result which covers the case of pairs of obstacles
with almost the same scattering length spectrum.

We refer the reader to Ch. 1 in \cite{PS} for the definition of generalized geodesics in the present (Euclidean) case, and to  
\cite{MS1}, \cite{MS2} or Sect. 24.3 \cite{H} for the definition and their main properties in more general situations. 

Let $K$ be an obstacle in $\R^n$ ($n\geq 2$) as in Sect. 1 and let
$$\tOm_K = \overline{{\R}^n\setminus K} .$$ 
 Given $\omega, \theta \in \sn$, a generalized geodesic $\gamma$ in $\tOm_K$ will be called an {\it $\ot$-ray} in
 $\tOm_K$ if $\gamma$ is unbounded in both directions, $\omega\in \sn$ is its incoming 
direction and $\theta\in \sn$ is its outgoing direction.
 
Next, we define the so called scattering length spectrum associated with an obstacle.

Let again $M$ be a large ball in $\R^n$ containing the obstacle $K$ in its interior and let $\bS_0 = \partial M$.
Given $\xi\in \sn$ denote by $Z_{\xi}$  { the hyperplane in ${\R}^n$ orthogonal to $\xi$ and tangent to $\bS_0$} such that $M$ is 
contained in the half-space $R_{\xi}$ determined by  $Z_{\xi}$ and having $\xi$ as an inner 
normal. For an $\ot$-ray $\gamma$ in $\Omega_K$, the {\it sojourn time} $T_{\gamma}$ 
of $\gamma$ is defined by $T_{\gamma} = T'_{\gamma} - 2a$, where $T'_{\gamma}$ is the length of 
that part of $\gamma$ which is contained in $R_{\omega}\cap R_{-\theta}$ and $a$ is the radius of 
the ball $M$. It is known (cf. \cite{G}) that this definition does not depend on the choice of  the ball $M$.  

 The {\it scattering length spectrum} of $K$ is defined to be the family 
of sets of real numbers $\sl_K =  \{ \sl_K\ot\}_{\ot}$ where $\ot$ runs over $\ssn$ and
$\sl_K\ot$ is the set of sojourn times $T_\gamma$ of all $\ot$-rays $\gamma$ in $\Omega_K$. 
It is known  (cf. \cite{PS}) that for $n \geq 3$, $n$ odd, and $C^\infty$ boundary $\dk$, we have 
$\sl_K\ot = \mbox{sing supp } s_K(t, \theta, \omega)$
for almost all $\ot$. Here $s_K$ is the {\it scattering kernel} related to the scattering operator for 
the wave equation in $\R\times \Omega_K$ with Dirichlet boundary condition on 
$\R\times \partial \Omega_K$ (cf. \cite{LP}, \cite{M}). Following \cite{St3}, 
we will say that two obstacles  $K$ and $L$ have {\it almost the same SLS} if there exists a subset 
$\rr$ of full Lebesgue measure in $\ssn$  such that  $\sl_K\ot = \sl_L\ot$ for all $\ot\in \rr$.

The flow $\fk_t$ can be made continuous using certain natural identifications of points at the boundary.
Consider the quotient space $T_b^*(\tOm_K) = T^*(\tOm_K)/\sim$ with respect to the equivalence relation: $(x,\xi) \sim (y,\eta)$
iff $x = y$ and either $\xi = \eta$ or $\xi$ and $\eta$ are symmetric with respect to the tangent plane to $\dk$ at $x$. Let
$\bS_b^*(\tOm_K)$ be the image of the {\it unit co-sphere bundle}  $\bS^*(\tOm_K)$. We will identify $T^*(\tOm_K)$ and 
$\bS^*(\tOm_K)$ with their images in $T_b^*(\tOm_K)$. 
It is known that for $K\in \kk$ the flow $\fk_t$ is well-defined and continuous (\cite{MS2}).
Some further regularity properties are established in \cite{St1} (see also Ch. 11 in \cite{PS}).

\bs

\noindent
{\bf Definition 2.1.} Let $K, L$ be two obstacles  in $\R^n$.
We will say that $\Omega_K$ and $\Omega_L$  {\it have conjugate flows} if  there exists a homeomorphism 
$$\Phi : \dt^*(\Omega_{K})\setminus \trapk  \longrightarrow  \dt^*(\Omega_{L})\setminus\trapl$$
which defines a symplectic map on an open dense subset of  $\dt^*(\Omega_{K})\setminus \trapk$,
it maps $\bS^*(\Omega_K)\setminus \trapk$ onto $\bS^*(\Omega_L)\setminus \trapl$,
and satisfies $\fl_t\circ \Phi = \Phi\circ \fk_t$ for all $t\in \R$ and $\Phi = \mbox{id}$ on 
$\dt^*(\R^n\setminus M)\setminus \trapk = \dt^*(\R^n\setminus M)\setminus\trapl$.

\bs


The following theorem was proved in  \cite{St3} in the case of the scattering length spectrum, and 
and then  in \cite{NS2} similar arguments were used to derive the case involving travelling times.

\bs

\noindent
{\bf Theorem 2.2.} { \it If the obstacles $K, L\in \kk_0$ have almost the same scattering length spectrum or almost the same
travelling times,  then $\Omega_K$ and $\Omega_L$ have conjugate flows.}

\bs

We can now state the main result in this paper.

\bs

\noindent
{\bf Theorem 2.3.} { \it Let the obstacles $K, L\in \kk_0$ satisfy the conditions {\rm (A0), (A1)} and {\rm  (A2)}.
If $\Omega_K$ and $\Omega_L$ have conjugate flows then $K = L$.}

\bs

Clearly, Theorem 1.1 is an immediate consequence of Theorem 2.3. We prove the latter in Sect. 3.

Form now on we will {\bf assume that the obstacles $K$ and $L$ in $M$ belong to the class $\kk_0$ and that $\Omega_K$ 
and $\Omega_L$ have conjugate flows}, that is there exists a homeomorphism $\Phi$ with the properties in Definition 2.1.

Next, we describe some propositions from \cite{St1}, \cite{St2} and \cite{St3} that are needed in the proof
of Theorem 2.3.



\bs

\noindent
{\bf Proposition 2.4.} (\cite{St1}, \cite{St2})

(a) {\it There exists a countable family $\{ M_i\} = \{ M_i^{(K)}\}$
of codimension $1$ submanifolds of $\bS^*_+(\bS_0)\setminus \trapk$ such that
every $\sigma \in \bS^*_+(\bS_0)\setminus (\trapk \cup_i M_i)$ generates
a simply reflecting ray in $\Omega_K$. Moreover the family  $\{ M_i\}$
is locally finite, that is any compact subset of  $\bS^*_+(\bS_0)\setminus \trapk$ has common points with
only finitely many of the submanifolds $M_i$.} 

\ms



(b)  {\it There exists a countable locally finite family $\{ P_i\}$
of codimension $2$ smooth submanifolds of $\bS^*_+(\bS_0)$ such that for any
$\sigma\in \bS^*_+(\bS_0)\setminus (\cup_i P_i)$ the trajectory $\gk(\sigma)$ contains no gliding segments on the boundary 
$\dk$ and $\gk(\sigma)$ contains at most one tangent point to $\dk$.}

\bs

It follows from the conjugacy of flows and Proposition 4.3 in \cite{St3} that
the submanifolds $M_i$ are the same for $K$ and $L$, i.e. $M_i^{(K)} = M_i^{(L)}$ for all $i$.
Notice that different submanifolds $M_i$ and $M_j$ may have common points
(these generate rays with more than one tangency to $\dk$) and in general are
not transversal to each other. However, as we see from part (b), if $M_i \neq M_j$
and $\sigma\in M_i \cap M_j$, then locally near $\sigma$, $M_i \neq M_j$, i.e. there exist points in
$M_i\setminus M_j$ arbitrarily close to $\sigma$.

From now on we will {\bf assume that $K$ and $L$ satisfy the condition (A1)} as well, apart from (A0).

Since $\bS^*_+(\bS_0)$ is a manifold and $\trapk$ and $\trapl$ are compact,  using the condition (A1) for 
both $K$  and $L$, it follows that any two points in $\bS^*_+(\bS_0)\setminus \trapk$ can be connected by a $C^1$ curve lying entirely 
in $\bS^*_+(\bS_0)\setminus (\trapk \cup \trapl)$.

Let  $\Gamma_K$ be the {\it set of the points $\sigma\in \bS^*_+(\bS_0)\setminus \trapk$
such that $\gamma_K(\sigma)$ is a simply reflecting ray}. It follows from \cite{MS2} (cf. also
Sect. 24.3 in \cite{H}) and Proposition 2.4 in \cite{St1}) that $\Gamma_K$ is open and dense
and has full Lebesgue measure in $\bS^*_+(\bS_0)$. Moreover, since $K,L$ have conjugate flows, 
Proposition 4.3 in \cite{St3} implies $\Gamma_K = \Gamma_L$. Finally, Proposition 6.3 in  \cite{St3} and the 
condition (A1) yield the following.
 
\bs

\noindent
{\bf Proposition 2.5.} {\it Let $K, L$ satisfy the conditions {\rm (A0)} and {\rm (A1)}. Then
\begin{equation}
\#(\gamma_K(\sigma) \cap \dk) = \# (\gamma_L(\sigma)\cap \dl) 
\end{equation}
for all $\sigma\in \Gamma_K = \Gamma_L$.}

\bs

That is, for $\sigma\in \Gamma_K = \Gamma_L$  the number of reflection points of 
$\gamma_K(\sigma)$ and $\gamma_L(\sigma)$ is the same.

\def\trho{\tilde{\rho}}

\section{Proof of Theorem 2.3}
\renewcommand{\theequation}{\arabic{section}.\arabic{equation}}
 
Let $K$ and $L$ be as in Theorem 2.3. We will show that they coincide.

Using the condition (A2) for $K$ and $L$, it follows that there exists a finite or countable family 
$\{Q_i\} $ of codimension two $C^1$ submanifolds of $\bS^*_+(\bS_0)\setminus (\trapk \cup \trapl)$  
which is locally finite in  $\bS^*_+(\bS_0)\setminus (\trapk \cup \trapl)$ and  such that for every 
$\rho \in \bS^*_+(\bS_0)\setminus (\trapk \cup  \trapl \cup  \cup_i Q_i)$ the billiard trajectory
$\gamma_K(\rho)$ has no conjugate points that both belong to $\dk$, and also the billiard trajectory
$\gamma_L(\rho)$ has no conjugate points that both belong to $\dl$.

Fix a  family $\{Q_i\} $ with this property. Fix also a countable family $\{M_i\}$ of codimension
one submanifolds of $\bS^*_+(\bS_0)$ with the property in Proposition 2.4(a) and  a countable family $\{ P_i\}$
of codimension $2$ smooth submanifolds of $\bS^*_+(\bS_0)$ having the property in Proposition 2.4(b)
for both $K$ and $L$.

We will use the general framework of the argument in Sect. 3 in \cite{NS1}.
Naturally, various modifications will be necessary.

As in \cite{NS1},  a point $y \in \dk$ will be called {\it regular} if $\dk = \dl$ in an open neighbourhood
of $y$ in $\dk$. Otherwise $y$ will be called {\it irregular}. The following definition is similar to the one 
in Sect. 7 in \cite{St3}.

\bs

\noindent
{\bf Definition.} A $C^1$ path $\sigma(s)$, $0\leq s\leq a$ (for some $a > 0$), in $\bS^*_+(\bS_0)\setminus (\trapk \cup \trapl)$  
will be called {\it admissible} if it has the following properties:

\begin{enumerate}

\item[(a)] $\sigma(0)$ generates a {\it free ray} in $\Omega_K$ and in $\Omega_L$,
i.e. a  ray without  any common points with $\dk$ and $\dl$.

\item[(b)]  if  $\sigma(s)\in M_i$ for some $i$ and $s\in [0,a]$, then $\sigma$ is
transversal to $M_i$ at $\sigma(s)$ and $\sigma(s) \notin M_j$ for any 
submanifold $M_j\neq M_i$ .

\item[(c)]  $\sigma(s)$ does not belong to any of the submanifolds $P_i$ and to any of the submanifolds $Q_i$
for all $s \in [0,a]$. 

\end{enumerate}

\bs

It follows from Proposition 6.3 in \cite{St3} (and its proof) that for every 
$$\rho \in \bS^*_+(\bS_0)\setminus (\trapk \cup \trapl \cup \cup_i P_i \cup \cup_i Q_i)$$ 
which belongs to at most one of the submanifolds $M_i$
there exists an admissible path   $\sigma(s)$, $0 \leq s\leq a$, with $\sigma(a) = \rho$.

Let $m \geq 1$ be an integer. As in \cite{NS1},
denote $Z_m$ be {\it the set of irregular points $x\in \dk$ with the following property:}
there exists an admissible path $\sigma(s)$, $0\leq s\leq a$, in $\bS^*_+(\bS_0)\setminus \trapk$ such that
$\sigma(a)$ generates a free ray in $\R^n$, $x$ belongs to the billiard trajectory
$\gk^+(\sigma(a))$ and for any $s\in [0,a]$ the trajectory $\gk^+(\sigma(s))$ has at
most $m$ irregular common points\footnote{And possibly a number of regular common points with $\dk$.} with $\dk$.

We will prove by induction on $m$ that $Z_m = \e$ for all $m \geq 1$.

\ms

\noindent
{\bf Step 1.} $Z_1 = \e$.  The proof of this case is the same as the one in \cite{NS1}. 
We sketch it here for completeness.
Assume that $Z_1 \neq \e$. Consider an arbitrary admissible path  $\sigma(s)$, $0\leq s\leq a$, in $\bS^*_+(\bS_0)$ such that
$\gk^+(\sigma(a))$ contains a point of $Z_1$  and for each  $s\in [0,a]$ the trajectory $\gk^+(\sigma (s))$ has at most $1$ 
irregular point. Take the minimal $a > 0$ with this property; then for any $s\in [0,a)$ the trajectory
$\gk^+(\sigma (s))$ contains no irregular points. Set $\rho = \sigma(a)$. It follows (as in \cite{NS1}) 
that $\gamma = \gk^+(\sigma(a))$ and  $\gamma' = \gl^+(\rho)$ have 
the same number of common points with $\dk$ and $\dl$, respectively.

Let $x_1, \ldots, x_k$ be the  common points of $\gk(\rho)$ with $\dk$ and let $x_i \in Z_1$ for some $i$. 
Then all $x_j$ with  $j\neq i$ are regular points, so there exists an open neighbourhood $U_j$ of $x_j$ in $\dk$ 
such that $U_j = U_j \cap \dl$. Setting $\rho = (x_0,u_0)$, let $t_{k+1} > 0$ be the largest number
such that $\rho_{k+1} = \fk_{t_{k+1}}(\rho) \in \bS^*(\bS_0)$. Let $\fk_{t_j}(\rho) = (x_j,u_j)$, $1 \leq j \leq k+1$.
It then follows from the above that
$\fk_t(\rho) = \fl_t(\rho)$ for $0 \leq t < t_i$, and also $\fk_{\tau}(\rho_{k+1}) = \fl_{\tau}(\rho_{k+1})$
for all $- (t_{k+1} - t_i) < \tau \leq 0$. So, the trajectories $\gk(\rho)$ and $\gl(\rho)$ both pass
through $x_{i-1}$ with the same (reflected) direction $u_{i-1}$ and through $x_{i+1}$ with the same
(reflected) direction $u_{i+1}$. Thus, $x_1, \ldots, x_{i-1}, x_{i+1}, \ldots, x_k$ are common points of
$\gl^+(\rho)$ and $\dl$. As observed above,  $\gl^+(\rho)$ must have exactly $k$ common points 
with $\dl$, so it has a common point $y_i$ with $\dl$ `between' $x_{i-1}$ and $x_{i+1}$.

Next, we consider two cases.

\ms

\noindent
{\bf Case 1.} $x_i$ is a transversal reflection point of $\gamma$ at $\dk$. Then the above shows that
$\gamma'$ has a transversal reflection point at $x_i$, so in particular $x_i \in \dl$. It is also clear that
for any $y\in \dk$  sufficiently close to $x_i$ there exists  $\rho'\in \bS^*_+(\bS_0)\setminus \trapk$ close to $\rho$ so that 
$\gk^+(\rho')$ has a proper reflection point at $y$, and repeating the previous argument and using again the fact that
$U_j = U_j \cap \dl$ for  $j \neq i$, we derive that  $y \in \dl$. Hence $\dk = \dl$ in an open neighbourhood of $x = x_i$ in $\dk$, 
which is impossible since $x$ is an irregular point.

\ms

\noindent
{\bf Case 2.} $x_i$ is a tangent point of $\gamma$ to $\dk$. Then each of the trajectories $\gamma$ and $\gamma'$ has 
exactly $k -1$ transversal reflection points. Moreover,  $x_i$ lies on the segment $[x_{i-1}, x_{i+1}]$. The trajectory
$\gamma'$ also has exactly one tangent point to $\dl$ and it must be a point $y_i$ on the segment $[x_{i-1}, x_{i+1}]$.
Assume for a moment that $y_i \neq x_i$. Then we can choose $x'_i \in \dk$ arbitrarily close to $x_i$ and 
$u'_i\in \sn$ close to $u_i$ so that $u'_i$ is tangent to $\dk$ at $x'_i$ and the straight line determined by $x'_i$
and $u'_i$ intersects $\dl$ transversally near $y_i$. Let $\rho'\in \bS^*_+(\bS_0)\setminus \trapk$ be the point close
to $\rho$ which determines a trajectory $\gk^+(\rho')$ passing through $x'_i$ in direction $u'_i$, i.e. tangent
to $\dk$ at $x'_i$. Then $\gk^+(\rho')$ has $k -1$ transversal reflections at $\dk$ and one tangent point, while 
$\gl^+(\rho')$ has $k$ transversal reflections at $\dl$ and no tangent points at all. This impossible, so we must have
$y_i = x_i$. A similar argument shows that every $x' \in \dk$ sufficiently close to $x_i$
belongs to $\dl$, as well. So, $x_i$ is a regular point, a contradiction. 

\ms

Thus we must have $Z_1 = \e$.

\ms

\noindent
{\bf Step 2: Inductive Step.} 
 Assume that  $Z_1 = \ldots = Z_{m-1} = \e$ for some integer $m > 1$. 
Suppose $Z_m \neq \e$. 
Then there exists an admissible  $C^1$ path $\sigma(s)$, $0\leq s\leq a$, in $\bS^*_+(\bS_0)$ 
such that $\gk^+(\sigma(a))$ has a common point with $Z_m$ and for each $s\in [0,a]$ the trajectory 
$\gk(\sigma (s))$ has at most $m$ irregular points. 
We may assume that $a > 0$ is minimal with this property. Then for $s \in [0,a)$, 
$\gk(\sigma (s))$  contains no points of $Z_m$, so if it passes through any irregular points, they
must be from some $Z_i $ with $i < m$. However  $Z_1 = \ldots = Z_{m-1} = \e$ by assumption,
so $\gk(\sigma (s))$  contains no irregular points at all for all $s \in [0,a)$. This implies
$\fk_t (\sigma(s)) = \fl_t(\sigma(s))$ for all $t\geq 0$ and all $s \in [0,a)$, and by continuity of the flows, we
derive $\fk_t (\sigma(a)) = \fl_t(\sigma(a))$ for all $t\geq 0$.

Set $\rho = \sigma(a)$, $\gamma = \gk^+(\rho)$, $\gamma' = \gl^+(\rho)$. It follows from the definition of
$Z_m$ and $\gamma \cap Z_m \neq \e$, that $\gamma$ contains at most $m$ irregular points.
Since $\gk^+(\sigma(s))$ has no irregular points at all for $s <a$, it follows that $\gamma$ contains
exactly $m$ irregular points; otherwise all irregular points in $\gamma$ would be in $Z_i$ for some $i < m$,
which is impossible since $Z_i = \e$.

Let $x_1, \ldots, x_m$ be the consecutive irregular common points of $\gamma$ with $\dk$, and let $y_1, \ldots, y_p$ be its
regular common points with $\dk$ (if any, i.e. we may have $p = 0$). Then from the definition of  a regular point, for each
$i = 1, \ldots,p$, there exists an open neighbourhood $V_i$ of $y_i$ in $\dk$ with $V_i \subset \dl$, i.e. $V_i$ is
an open neighbourhood of $y_i$ in $\dl$ as well. Since $\sigma(s)$ is an admissible path, if $\gamma$ has a tangent point to $\dk$,
then it is exactly one of the points $x_1, \ldots, x_m, y_1, \ldots, y_p$. 
Let $\pr_1(\fk_{\tau_i}) = y_i = \pr_1(\fl_{\tau_i})$ for some $0 < \tau_1 < \ldots < \tau_p$. It follows from the continuity of the flows
$\fk_t$ and $\fl_t$ that there exists an open neighbourhood $W$ of $\rho$ in $\bS^*_+(\bS_0)$ and $\delta > 0$ such that for any
$\rho' \in W$ if  $\fk_\tau(\rho') \in \dk$ (or  $\fl_\tau(\rho') \in \dl$) for some $\tau$ with $|\tau - \tau_i| < \delta$, then 
$\pr_1(\fk_\tau(\rho')) \in V_i$ (resp. $\pr_1(\fl_\tau(\rho')) \in V_i$). In particular, taking $W$ sufficiently small we have that
$\gk^+(\rho')$ contains at most $m$ irregular points for all $\rho' \in W$.

Since $\sigma(s)$ is an admissible path, is does not contain any points from $\cup_i Q_i$. In particular,
$\rho \notin \cup_i Q_i$, so $\gamma$ does not have conjugate points both belonging to $\dk$.

Let $\pr_1(\fk_{t_j}(\rho)) = x_j = \pr_1(\fl_{t_j}(\rho))$, $ j = 1, \ldots, m$, for some $0 < t_1 < \ldots < t_m$, and let
$u_j = \pr_2(\fk_{t_j^+}(\rho))$ be the reflected direction of the trajectory $\gamma$ at $x_j$.

\ms

\noindent
{\bf Case 1.}  The points $x_1, \ldots, x_m$ are all transversal reflection points of $\gamma$ (some of the points
$y_i$ might be a tangent points of $\gamma$ to $\dk$).  Then for $s < a$ close to $a$, $\gk^+(\sigma(s))$ has
transversal reflection points $x_1(s), \ldots, x_m(s)$ close to $x_1, \ldots, x_m$, respectively, that is $x_i(s) \to x_i$
as $ s\nearrow a$.  Since $\gk^+(\sigma(s))$ has only regular reflection points, we have $\dk = \dl$ in an open 
neighbourhood of $x_i(s)$ in $\dk$ for $s< a$ close to $a$. Hence  there exists an open subset $U_i$ of $\dk$ with
$x_i \in \overline{U_i}$ and $\dk \cap U_i = \dl \cap U_i$ for all $i = 1, \ldots, m$.

Set $\rho_j = \fk_{t_j}(\rho)$.  
Take a small number $\delta > 0$ and consider 
$$\oo =  \{ u \in \sn : \|u - u_1\| < \delta\} .$$
If $\delta$ is small enough, for every $u \in \oo$ there exists $t = t(u) \in \R$ close to $t_2-t_1$ such that
$G(u) = \pr_1 (\fk_{t(u)}(\rho_1)) \in \dk$. This defines a smooth local map $G : \oo \longrightarrow \dk$
taking values near $x_2$. Since $\gamma$ has no conjugate points both belonging to $\dk$, $x_1$ and $x_2$
are not conjugate points along $\gamma$. This implies that the linear map $dG(u_1)$ has full rank $n-1$, and therefore
$G(\oo)$ covers a whole open neighbourhood of $x_2$ in $\dk$. Since $x_2 \in \overline{U_2}$, it follows that
there exist $u \in \oo$ arbitrarily close to $u_1$ such that $G(u) \in U_2$. Given such an $u \in \oo$, consider the billiard 
trajectory in $\Omega_K$ generated by $(x_1,u)$, and let $\rho'\in \bS^*_+(\bS_0)$ be the point that belongs to this billiard trajectory.
Then $\gk^+(\rho')$ has transversal reflection points $x'_1,x'_2, \ldots, x'_m$ close to the points $x_1,x_2, \ldots,x_m$,
respectively. 
If $u$ is sufficiently close to $u_1$, then $\rho' \in W$, so $\gk^+(\rho')$ contains at most $m$ irregular points.
On the other hand, for such $u$ we have $x'_2 = G(u) \in U_2$ and we have $\dk = \dl$ on $U_2$. Thus, $x'_2$ cannot be
an irregular point. There are no other places from which $\gk^+(\rho')$ can gain an irregular point, therefore it turns out
$\gk^+(\rho')$ has at most $m-1$ irregular points. Now the assumption $Z_1 = Z_2 = \ldots = Z_{m-1} = \e$ gives that
$\gk^+(\rho')$ contains no irregular points at all. However $x_1$ is one of the reflection points of $\gk^+(\rho')$ 
and $x_1$ is irregular, a contradiction. 

This proves that Case 1 is impossible.

\ms

\noindent
{\bf Case 2.}  One of the points $x_1, \ldots, x_m$ is a tangent point of $\gamma$ to $\dk$. Then all other $x_i$ are
transversal reflections, and since $m > 1$, we have at least one such point. We will assume that one of the points $x_1$ or $x_2$
is a tangent point of $\gamma$ to $\dk$; the other cases are considered similarly.
Assume e.g. that $\gamma$ is tangent to $\dk$ at $x_2$ and has a transversal reflection at $x_1$; otherwise we will change the 
roles of $x_1$ and $x_2$  and reverse the motion along the trajectory.  
Since $\gamma$ can have only one tangent point to $\dk$, all points
$y_i$ (if any) are transversal reflection points. Moreover we have $\dk = \dl$ in an open neighbourhood $V_i$ of
$y_i$ in $\dk$. Take again a small number $\delta > 0$ and consider $\oo =  \{ u \in \sn : \|u - u_1\| < \delta\}$.
Given $u \in \oo$, consider the billiard  trajectory in $\Omega_K$ generated by $(x_1,u)$, and let 
$\rho(u)\in \bS^*_+(\bS_0)$ be the point that belongs to this billiard trajectory. Assuming that $\delta$ is small enough,
for all $u \in \oo$ the trajectory $\gk^+(\rho(u))$ has transversal reflection points $x_j(u)$ close to $x_j$
for $j \neq 2$ and $y_i(u)$ close to $y_i$ for all $i$. It may not have a common point with $\dk$ near $x_2$.
In fact, we will show that we can choose $u \in \oo$ arbitrarily close to $u_1$ so that $\gk^+(\rho(u))$ has 
no common point with $\dk$ near $x_2$. This is obvious when there are no reflection points of $\gamma$ between
$x_1$ and $x_2$, so assume that some of the reflections at the points $y_i$ occur between $x_1$ and $x_2$.
Let $y_i$ be the last reflection point of $\gamma$ before $x_2$, i.e. $t_1 < \tau_i < t_2 < \tau_{i+1}$. Recall
the open neighbourhood $V_i$ of $y_i$ in $\dk$ with $\dk = \dl$ on $V_i$ (i.e. $V_i \subset \dl$). 

Denote by $\Sigma$ the hyperplane in $\R^n$ passing through $x_2$ and perpendicular to $u_2$. Define the
map $G: \oo \longrightarrow \Sigma$ as follows: given $u\in \oo$, the trajectory $\gk^+(\rho(u))$ reflects
at $y_i(u)$ on $V_i$ with a reflected direction $\eta_i(u)$, and the straight-line ray issued from $y_i(u)$ in
direction $\eta_i(u)$ intersects $\Sigma$ at some point  which we call $G(u)$. It is clear that $G$ is a smooth map
(only transversal reflections occur between $x_1$ and $x_2$) and $G(u_1) = x_2$. Since $\gamma$ does not
have conjugate points both belonging to $\dk$, we have $\rank (dG(u_1)) = n-1$, so $G(\oo)$ contains a whole
open neighbourhood $V$ of $x_2$ in $\Sigma$. Assuming that the neighbourhood $V$ of $x_2$ is
sufficiently small, $V \setminus K$ contains a non-trivial open subset whose closure contains $x_2$. Thus,
there exist $u \in \oo$ arbitrarily close to $u_1$ for which $G(u) \in V \setminus K$, which means that the trajectory
$\gk^+(\rho(u))$ will intersect $\Sigma$ in $V \setminus K$ and so it will not have a common point with $\dk$
near $x_2$. In particular, $\gk^+(\rho(u))$ will have at most $m -1$ irregular points, and the assumption now yields
that $\gk^+(\rho(u))$ has no  irregular points at all. This is a contradiction, since $x_1$ is an irregular point
and it belongs to $\gk^+(\rho(u))$.

Hence Case 2 is impossible as well. This proves that we must have $Z_m = \e$. 

\ms

By induction  $Z_m = \e$ for all $m \geq 1$, so there are no irregular points at all.

It is now easy to prove that $\dk \subset \dl$. Let $A$ be the set of those
$$\rho \in  \bS^*_+(\bS_0)\setminus (\trapk \cup \trapl \cup \cup_i P_i \cup \cup_i Q_i) .$$ 
such that $\rho$ belongs to at most one of the submanifolds $M_i$. Clearly, $A$ is a
dense subset of $\bS^*_+(\bS_0)$. Moreover, the set $B$ of the points $x\in \dk$ that belong to 
$\gk^+(\rho)$ for some $\rho \in A$  is dense in $\dk$.  As we mentioned earlier,
it follows from Proposition 6.3 in \cite{St3} (and its proof) that for every  $\rho \in A$
there exists an admissible path  $\sigma(s)$, $0 \leq s\leq a$, with $\sigma(a) = \rho$.
Since $Z_m = \e$ for all $m$, we derive $\gk^+(\rho) = \gl^+(\rho)$. 
The latter is then true for all $\rho \in A$, and therefore $B \subset \dl$. Since $B$ is
dense in $\dk$, it follows that $\dk \subset \dl$.

By symmetry we get $\dl \subset \dk$ as well, so $\dk = \dl$.
\endofproof

\bs

\ms

\noindent
School of Mathematics and Statistics, University of Western Australia, Crawley 6009 WA, Australia\\
E-mail: luchezar.stoyanov@uwa.edu.au

\end{document}